\documentclass[prc,superscriptaddress,showpacs,twocolumn,amssymb,amsmath,amsfonts,aps,floatfix,letterpaper]{revtex4}
\usepackage{graphicx} 
\usepackage{dcolumn}  
\usepackage{longtable}
\usepackage{bm}       
\usepackage{subfigure}
\newcommand\captionof[1]{\def\@captype{#1}\caption}
\begin{document}
%
%
%
%
\newcommand*{\CMU}{Carnegie Mellon University, Pittsburgh, Pennsylvania 15213}
\newcommand*{\CMUindex}{1}
\affiliation{\CMU}
\newcommand*{\ANL}{Argonne National Laboratory, Argonne, Illinois 60441}
\newcommand*{\ANLindex}{2}
\affiliation{\ANL}
\newcommand*{\ASU}{Arizona State University, Tempe, Arizona 85287-1504}
\newcommand*{\ASUindex}{3}
\affiliation{\ASU}
\newcommand*{\CSU}{California State University, Dominguez Hills, Carson, CA 90747}
\newcommand*{\CSUindex}{4}
\affiliation{\CSU}
\newcommand*{\Canisius}{Canisius College, Buffalo, NY 14208}
\newcommand*{\Canisiusindex}{5}
\affiliation{\Canisius}
\newcommand*{\CUA}{Catholic University of America, Washington, D.C. 20064}
\newcommand*{\CUAindex}{6}
\affiliation{\CUA}
\newcommand*{\SACLAY}{CEA, Centre de Saclay, Irfu/Service de Physique Nucl\'eaire, 91191 Gif-sur-Yvette, France}
\newcommand*{\SACLAYindex}{7}
\affiliation{\SACLAY}
\newcommand*{\CNU}{Christopher Newport University, Newport News, Virginia 23606}
\newcommand*{\CNUindex}{8}
\affiliation{\CNU}
\newcommand*{\UCONN}{University of Connecticut, Storrs, Connecticut 06269}
\newcommand*{\UCONNindex}{9}
\affiliation{\UCONN}
\newcommand*{\ECOSSEE}{Edinburgh University, Edinburgh EH9 3JZ, United Kingdom}
\newcommand*{\ECOSSEEindex}{10}
\affiliation{\ECOSSEE}
\newcommand*{\FU}{Fairfield University, Fairfield CT 06824}
\newcommand*{\FUindex}{11}
\affiliation{\FU}
\newcommand*{\FIU}{Florida International University, Miami, Florida 33199}
\newcommand*{\FIUindex}{12}
\affiliation{\FIU}
\newcommand*{\FSU}{Florida State University, Tallahassee, Florida 32306}
\newcommand*{\FSUindex}{13}
\affiliation{\FSU}
\newcommand*{\GWU}{The George Washington University, Washington, DC 20052}
\newcommand*{\GWUindex}{14}
\affiliation{\GWU}
\newcommand*{\ECOSSEG}{University of Glasgow, Glasgow G12 8QQ, United Kingdom}
\newcommand*{\ECOSSEGindex}{15}
\affiliation{\ECOSSEG}
\newcommand*{\ISU}{Idaho State University, Pocatello, Idaho 83209}
\newcommand*{\ISUindex}{16}
\affiliation{\ISU}
\newcommand*{\INFNFR}{INFN, Laboratori Nazionali di Frascati, 00044 Frascati, Italy}
\newcommand*{\INFNFRindex}{17}
\affiliation{\INFNFR}
\newcommand*{\INFNGE}{INFN, Sezione di Genova, 16146 Genova, Italy}
\newcommand*{\INFNGEindex}{18}
\affiliation{\INFNGE}
\newcommand*{\INFNRO}{INFN, Sezione di Roma Tor Vergata, 00133 Rome, Italy}
\newcommand*{\INFNROindex}{19}
\affiliation{\INFNRO}
\newcommand*{\ORSAY}{Institut de Physique Nucl\'eaire ORSAY, Orsay, France}
\newcommand*{\ORSAYindex}{20}
\affiliation{\ORSAY}
\newcommand*{\ITEP}{Institute of Theoretical and Experimental Physics, Moscow, 117259, Russia}
\newcommand*{\ITEPindex}{21}
\affiliation{\ITEP}
\newcommand*{\JMU}{James Madison University, Harrisonburg, Virginia 22807}
\newcommand*{\JMUindex}{22}
\affiliation{\JMU}
\newcommand*{\KYUNGPOOK}{Kyungpook National University, Daegu 702-701, Republic of Korea}
\newcommand*{\KYUNGPOOKindex}{23}
\affiliation{\KYUNGPOOK}
\newcommand*{\UNH}{University of New Hampshire, Durham, New Hampshire 03824-3568}
\newcommand*{\UNHindex}{24}
\affiliation{\UNH}
\newcommand*{\NSU}{Norfolk State University, Norfolk, Virginia 23504}
\newcommand*{\NSUindex}{25}
\affiliation{\NSU}
\newcommand*{\OHIOU}{Ohio University, Athens, Ohio  45701}
\newcommand*{\OHIOUindex}{26}
\affiliation{\OHIOU}
\newcommand*{\ODU}{Old Dominion University, Norfolk, Virginia 23529}
\newcommand*{\ODUindex}{27}
\affiliation{\ODU}
\newcommand*{\RPI}{Rensselaer Polytechnic Institute, Troy, New York 12180-3590}
\newcommand*{\RPIindex}{28}
\affiliation{\RPI}
\newcommand*{\URICH}{University of Richmond, Richmond, Virginia 23173}
\newcommand*{\URICHindex}{29}
\affiliation{\URICH}
\newcommand*{\ROMAII}{Universita' di Roma Tor Vergata, 00133 Rome Italy}
\newcommand*{\ROMAIIindex}{30}
\affiliation{\ROMAII}
\newcommand*{\MOSCOW}{Skobeltsyn Nuclear Physics Institute, Skobeltsyn Nuclear Physics Institute, 119899 Moscow, Russia}
\newcommand*{\MOSCOWindex}{31}
\affiliation{\MOSCOW}
\newcommand*{\SCAROLINA}{University of South Carolina, Columbia, South Carolina 29208}
\newcommand*{\SCAROLINAindex}{32}
\affiliation{\SCAROLINA}
\newcommand*{\JLAB}{Thomas Jefferson National Accelerator Facility, Newport News, Virginia 23606}
\newcommand*{\JLABindex}{33}
\affiliation{\JLAB}
\newcommand*{\UNIONC}{Union College, Schenectady, NY 12308}
\newcommand*{\UNIONCindex}{34}
\affiliation{\UNIONC}
\newcommand*{\UTFSM}{Universidad T\'{e}cnica Federico Santa Mar\'{i}a, Casilla 110-V Valpara\'{i}so, Chile}
\newcommand*{\UTFSMindex}{35}
\affiliation{\UTFSM}
\newcommand*{\VIRGINIA}{University of Virginia, Charlottesville, Virginia 22901}
\newcommand*{\VIRGINIAindex}{36}
\affiliation{\VIRGINIA}
\newcommand*{\WM}{College of William and Mary, Williamsburg, Virginia 23187-8795}
\newcommand*{\WMindex}{37}
\affiliation{\WM}
\newcommand*{\YEREVAN}{Yerevan Physics Institute, 375036 Yerevan, Armenia}
\newcommand*{\YEREVANindex}{38}
\affiliation{\YEREVAN}
 
\newcommand*{\NOWIMPERIAL}{Imperial College London, London SW7 2AZ, UK} 
\newcommand*{\NOWSTANFORD}{Stanford University, Stanford, CA 94305}
\newcommand*{\NOWMINN}{University of Minnisota, Minneapolis, MN 55455}

\newcommand*{\NOWLPSC}{LPSC-Grenoble, France}
\newcommand*{\NOWJLAB}{Thomas Jefferson National Accelerator Facility, Newport News, VA 23606}
\newcommand*{\NOWLANL}{Los Alamos National Laborotory, NM}
\newcommand*{\NOWCNU}{Christopher Newport University, Newport News, VA 23606}
\newcommand*{\NOWECOSSEE}{Edinburgh University, Edinburgh EH9 3JZ, UK}
\newcommand*{\NOWWM}{College of William and Mary, Williamsburg, VA 23187-8795}

\author{M.~Williams}
\altaffiliation[Current address: ]{\NOWIMPERIAL}
\affiliation{\CMU}
\author{Z.~Krahn}
\altaffiliation[Current address: ]{\NOWMINN}
\affiliation{\CMU}
\author {D.~Applegate}
\altaffiliation[Current address: ]{\NOWSTANFORD}
\affiliation{\CMU}
\author {M.~Bellis} 
\affiliation{\CMU}
\author {C.A.~Meyer}
\affiliation{\CMU}

\author {K. P. ~Adhikari} 
\affiliation{\ODU}
\author {M.~Anghinolfi} 
\affiliation{\INFNGE}
\author {H.~Baghdasaryan} 
\affiliation{\VIRGINIA}
\affiliation{\ODU}
\author {J.~Ball} 
\affiliation{\SACLAY}
\author {M.~Battaglieri} 
\affiliation{\INFNGE}
% ...

\collaboration{The CLAS Collaboration}
\noaffiliation


\date{\today}

\title{Differential cross sections for the reactions 
  $\gamma p \rightarrow p \eta$ and $\gamma p \rightarrow p \eta^{\prime}$}
%
%
\begin{abstract} 
High-statistics differential cross sections for the reactions 
$\gamma p \rightarrow p \eta$ and $\gamma p \rightarrow p \eta^{\prime}$ have 
been measured using the CLAS at Jefferson Lab for center-of-mass energies from near threshold
up to $2.84$~GeV. The $\eta^{\prime}$ results are the most precise to date and provide the 
largest energy and angular coverage.  The $\eta$ measurements extend the energy range 
of the world's large-angle results by approximately $300$~MeV. These new data, in particular
the $\eta^{\prime}$ measurements, are likely to help constrain the analyses being performed
to search for new baryon resonance states.
\end{abstract} 
\maketitle
\section{\label{section:intro}INTRODUCTION}

Studying low-energy $\eta$ and $\eta^{\prime}$ photoproduction presents an
interesting opportunity to search for new baryon resonances. 
Since both of these mesons have isospin $I=0$, the $N\eta$ and $N\eta^{\prime}$
final states couple to $N^*$ states but not $\Delta^*$ states.
Previous experiments have produced precise cross section measurements for the
${\gamma p\rightarrow p \eta}$ reaction from threshold up to a center-of-mass
energy, $W$, of approximately $2.5$~GeV~\cite{dugger-eta,crede,graal02,lns06}. 
For the ${\gamma p\rightarrow p \eta^{\prime}}$ reaction, previous results are 
fairly precise from threshold up to $W\approx 2.2$~GeV~\cite{dugger-etaprime}.

Studies performed on these previous experimental data have yielded evidence
for nucleon resonance contributions. For example, Anisovich 
{\em et al}.~\cite{anisovich} confirmed
that $\eta$ photoproduction is dominated near threshold by contributions from 
the $S_{11}(1535)$ and $S_{11}(1650)$ states. Evidence was 
also found for contributions from other resonances at higher energies, along 
with strong $t$-channel contributions in the forward direction.
The previously published CLAS $\eta^{\prime}$ results support contributions from
several resonance states as well~\cite{dugger-etaprime}.

The $\eta^{\prime}$ results presented in this paper are more precise than any
previous measurements and extend the large-angle energy range by approximately 
$600$~MeV. They will provide stronger constraints on models that attempt to 
extract resonance contributions in this reaction. Our $\eta$ measurements extend 
the energy range of the world's large-angle results by approximately $300$~MeV.
Significant discrepancies exist between our $\eta$ results and those previously 
published by CB-ELSA~\cite{crede} at higher energies (see Section~\ref{section:results}).  
These new results for both $\eta$ and $\eta^{\prime}$ will surely have an impact on 
the physics interpretation of these reactions. 

\section{\label{section:setup}EXPERIMENTAL SETUP}
The data were obtained using the CEBAF Large Acceptance Spectrometer (CLAS) 
housed in Hall B at the Thomas Jefferson National Accelerator Facility in
Newport News, Virginia. A 4~GeV electron beam hitting a $10^{-4}$ radiation length
gold foil produced real photons via the bremsstrahlung process. The recoiling 
electrons were then analyzed using a dipole magnet and scintillator hodoscopes 
in order to obtain, or ``tag'', the energy of the photons~\cite{sober}. Photons
in the energy range from $20\%$ to $95\%$ of the electron beam energy were tagged
and thus measured with an energy resolution of $0.1\%$ of the electron beam energy. 
The data were analyzed in center-of-mass energy bins that varied in width from 
$10$~MeV up to $40$~MeV, depending on the statistics.

The physics target, which was filled with liquid hydrogen, was a $40$-cm long
cylinder with a radius of $2$~cm. Continuous monitoring of the temperature and
pressure permitted determination of the density with uncertainty of $0.2\%$. 
The target cell was surrounded by $24$ ``start counter'' scintillators that were 
used in the event trigger. 

The CLAS detector utilized a non-uniform toroidal magnetic field of peak 
strength near $1.8$~T in conjunction with drift chamber tracking to determine 
particle momenta. The detector was divided into six sectors, such that when 
viewed along the beam line it was six-fold symmetric. Charged particles with laboratory 
polar angles in the range $8^{\circ}-140^{\circ}$ could be tracked over approximately 
$83$\% of the azimuthal angle. A set of $288$ scintillators placed outside of the magnetic 
field region was used in the event trigger and during off-line analysis in 
order to determine time of flight (TOF) of charged particles. The momentum resolution 
of the detector was, on average, about $0.5$\%. Other components of the 
CLAS, such as the Cerenkov counters 
and the electromagnetic calorimeters, were not used in this analysis. A more detailed 
description of the CLAS can be found in reference~\cite{clas-detector}.

The events were collected using a ``two-track'' trigger. This trigger required a coincidence 
between signals from the photon tagger and the CLAS. The signal from the tagger consisted 
of an OR of $40$ of the $61$ timing scintillators, corresponding to photon energies above 
$1.58$~GeV. This run setup where only part of the tagger was included in the trigger was
intended to mainly detect photons above $1.58$~GeV, avoiding the large bremsstrahlung
contribution from photons of lower energy. However, events originating from a photon
of energy below $1.58$~GeV (corresponding to a hit in counters $41$-$61$) were 
recorded when they had an accidental (random) hit in one of the triggered tagger counters. 
By using these ``accidental'' events, we were able to collect data for photons with energies
from $\sim 1$~GeV to $1.58$~GeV. This required applying an appropriate renormalization
based on the probability for such a random coincidence to occur (see Sec.~\ref{section:norm}). 
The signal from the CLAS required at least two sector-based signals. 
These signals consisted of an OR of any of the 4 start counter scintillators in 
coincidence with an OR of any of the $48$ time-of-flight scintillators in the sector. 
The rate at which hadronic events were accumulated was about $5$~kHz; however, only a 
small fraction of these events contained the reactions of interest to the analysis 
presented here.

\section{\label{section:data}Data and Event Selection}
The data reported here were obtained in the summer of 2004 during the CLAS ``g11a''
data taking period, in which approximately $20$ billion triggers were recorded. The 
relatively loose electronic trigger led to accumulation of data for a number of 
photoproduction reactions. During offline calibration, the timing of the photon 
tagger, the start counter and the time-of-flight elements were aligned with each other.
Calibrations were also made for the drift times of each of the drift chamber 
packages and the pulse heights of each of the time-of-flight counters. Finally, 
processing of the raw data was performed in order to reconstruct tracks in the 
drift chambers and match them with time-of-flight counter hits.

The reconstructed tracks were corrected for small imperfections in the magnetic 
field map and drift chamber alignment, along with their mean energy losses as
they passed through the target, the beam pipe, the start counter and air.
In addition, small corrections were made to the incident photon energies to
account for mechanical sag in the tagger hodoscope.

The CLAS was optimized for detection of charged particles; thus, the 
$\pi^{+}\pi^{-}\pi^{0}$ decay of the $\eta$ and the $\pi^{+}\pi^{-}\eta$ decay
of the $\eta^{\prime}$ were used to select the reactions of interest in this analysis. 
Detection of two positively charged particles and one negatively charged particle 
was required. A 1-constraint kinematic fit to the hypothesis 
$\gamma p \rightarrow p\pi^{+}\pi^{-}(\pi^{0}/\eta)$ was performed. This fit adjusts
the momenta of all measured particles within their measurement errors such that
energy and momentum are conserved and the missing mass is that of either a $\pi^{0}$
or an $\eta$. The shifts in the momentum, combined with the known errors, yields
a $\chi^{2}$  which is then converted to a
probability (confidence level) that the event is the desired reaction. A cut was 
placed on the resulting confidence levels in order to select events consistent with 
one of the two topologies. Fits were run for each of the possible $p,\pi^+$ particle identification 
assignments using each of the recorded photons in the event. Photon-particle combinations 
with confidence levels greater than $10\%$ were retained for further analysis. The trial
identity as a proton or a pion for positive particles (assigned by the kinematic fit) was
then checked using time-of-flight and momentum measurements.

\begin{figure}[h!]\centering
\subfigure[]{
  \label{confidence_level}
  \includegraphics[width=0.40\textwidth]{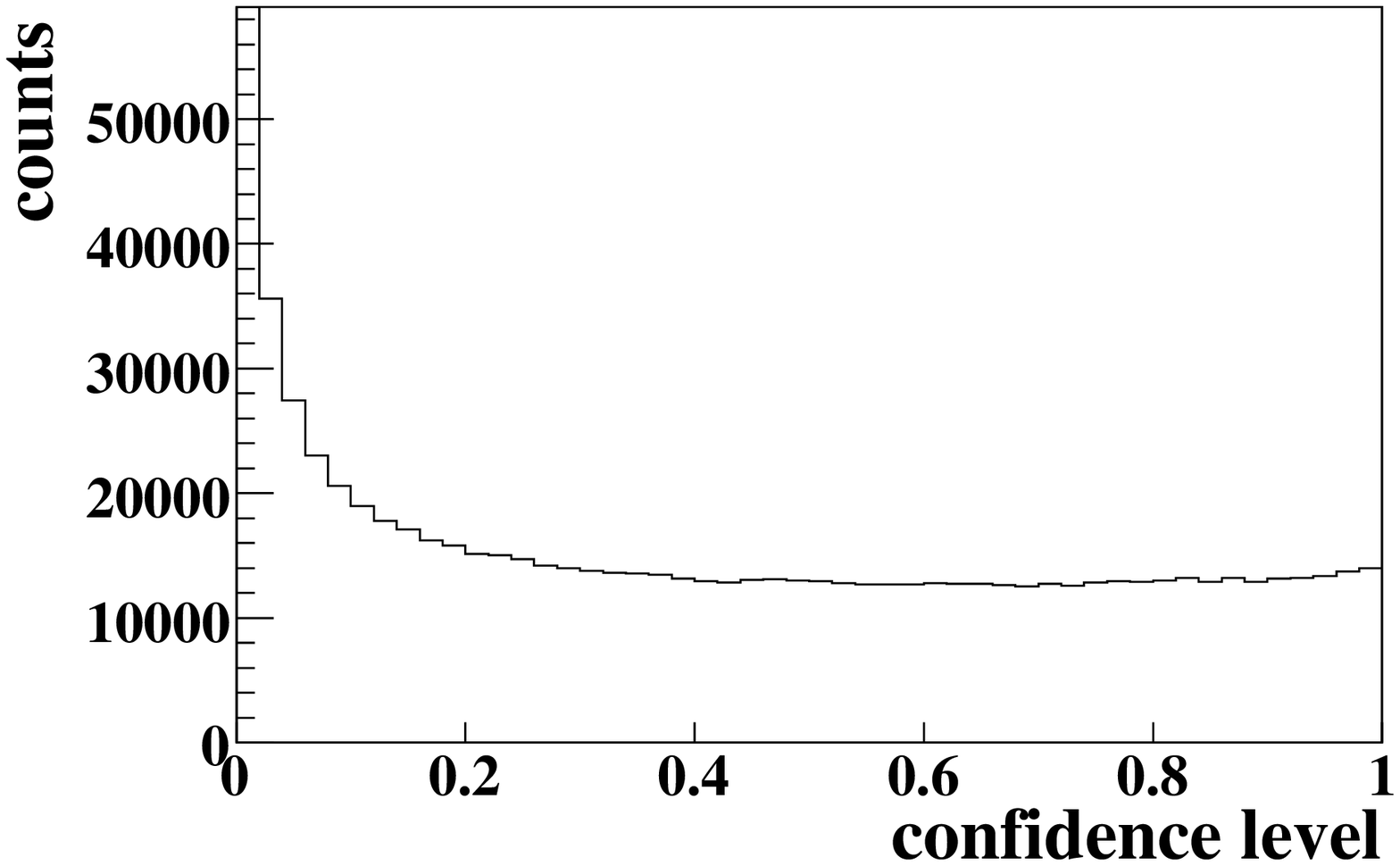}
} \\
\subfigure[]{
  \label{pull_dist}
  \includegraphics[width=0.40\textwidth]{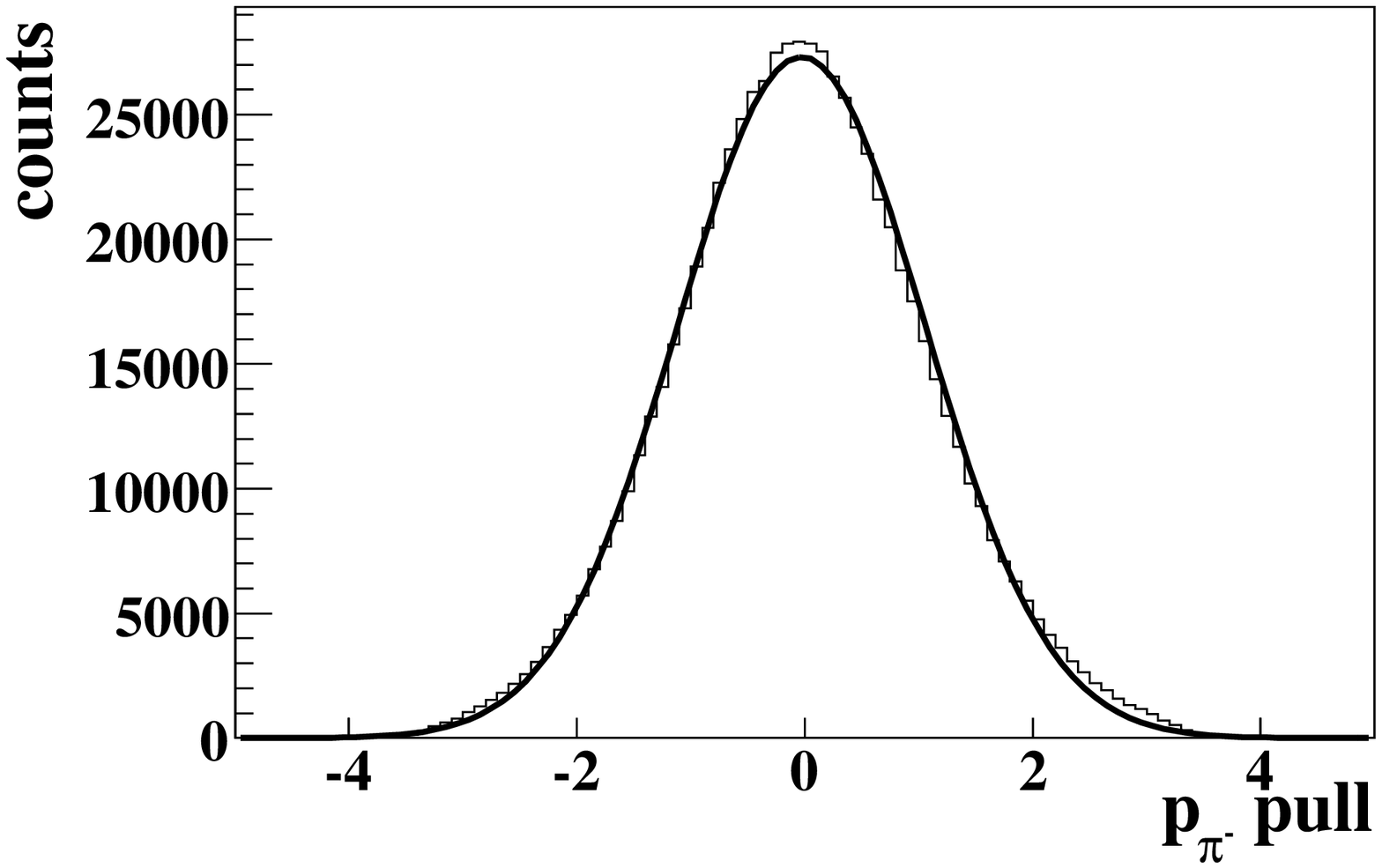}
}
\caption[]{\label{fig:kfit}
  (a) The confidence levels resulting from four-constraint kinematic fits
  performed on a sample of events to the calibration hypothesis
  $\gamma p \rightarrow p \pi^+ \pi^-$ integrated over all kinematics.
  The ``peak'' near zero consists of events that do not match the hypothesis,
  along with poorly measured (due to multiple scattering, \emph{etc}.)
  signal events. Agreement with the ideal (flat)
  distribution for signal events is very good.
  (b) Example pull-distribution for the momentum of the $\pi^{-}$ from the same
  kinematic fits as in (a).
  Only events with a confidence level larger than 1\% are shown. The line
  represents a Gaussian fit to this distribution. For this event sample, the
  parameters obtained are $\mu = -0.029\pm0.001,\sigma = 1.086\pm0.001$
  (the uncertainties are purely statistical), which
  are in very good agreement with the ideal values $\mu = 0,\sigma = 1$.
  Both (a) and (b) are good indicators that the CLAS error matrix is well
  understood. We note that the 10\% confidence-level cut used in the analysis
corresponds to the relatively flat region of the confidence-level plot. (This figure
is reproduced from reference~\cite{omega-prc}).
}
\end{figure}

The covariance matrix of the measured momenta was studied using four-constraint kinematic 
fits performed on the exclusive reaction $\gamma p \rightarrow p \pi^+ \pi^-$ in both 
real and Monte Carlo data samples. The confidence levels in all kinematic regions were 
found to be sufficiently flat and the pull-distributions (stretch functions) were 
Gaussians centered at zero with $\sigma = 1$ (see Fig.~\ref{fig:kfit}). The uncertainty 
in the extracted yields due to differences in signal lost because of this confidence-level cut in real 
as compared to Monte Carlo data is estimated to be about $3$\%.

All negatively charged tracks were assigned a $\pi^{-}$ identification. For positively charged 
tracks, the trial identification from the kinematic fit was checked using time-of-flight 
and momentum measurements. 
The tagger signal time, which was synchronized with the accelerator radio-frequency (RF) timing,
is used to obtain the start time for the event by accounting for the photon time of flight from
the tagger to the reaction vertex.
The stop time for each track was obtained from the TOF scintillator element hit by the 
track. The difference between these two times was the measured time of flight, $t_{meas}$. 
Track reconstruction through the CLAS magnetic field yielded both the momentum, $\vec{p}$, 
of each track, along with the path length, $L$, from the reaction vertex to the 
time-of-flight counter hit by the track. The expected time of flight for a mass hypothesis, 
$m$, is then given by
\begin{equation}
  t_{exp} = \frac{L}{c}\sqrt{1 + \left(\frac{m}{p}\right)^2}.
\end{equation}
The difference in these two time-of-flight calculations, 
$\Delta tof = t_{meas} - t_{exp}$, 
was used in order to separate protons from pions and to remove events associated with 
incorrect photons.

\begin{figure}[h]
  \centering
  \includegraphics[width=0.45\textwidth]{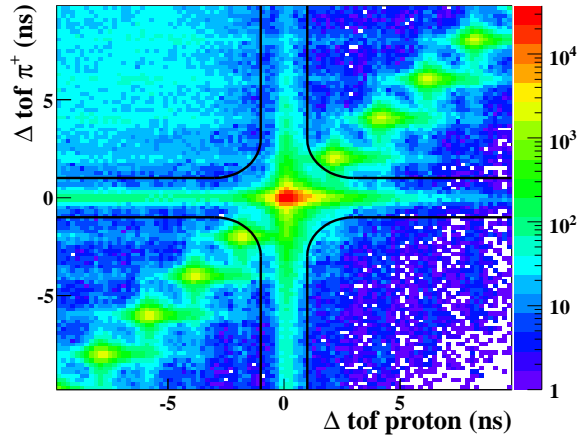}
\caption[]{\label{fig:pid}
  (Color Online)
  $\Delta tof_{\pi^+}$(ns) {\em vs.} $\Delta tof_{p}$(ns):  
  Particle identification cut for a sample of events that pass a 10\% 
  confidence level cut when kinematically fit to the hypothesis
  $\gamma p \rightarrow p \pi^+ \pi^- (\pi^0)$.
  The black lines indicate the timing cuts. Note the logarithmic scale on the
  intensity axis.
}
\end{figure}

Figure~\ref{fig:pid} shows $\Delta tof$ for tracks passing the kinematic fit 
under the $\pi^+$ hypothesis {\em vs.} $\Delta tof$ for the track passing the 
fit under the proton hypothesis. The region near $(0,0)$ contains events where
both tracks are good matches to their respective particle identification 
hypotheses. The $2$~ns radio-frequency time structure of the accelerator is evident 
in the out-of-time event clusters. Events outside of the black lines, where 
neither hypothesis was met, were cut from our analysis. This cut was designed 
to remove a minimal amount of good events. The Feldman-Cousins method \cite{feldman} was 
used to place an upper limit on the signal lost 
at $1.3\%$. Any remaining accidental events fell into the broad background under the 
$\eta/\eta^{\prime}$, and were rejected during the signal-background separation stage of 
the analysis discussed in Section~\ref{section:sig-bkgd}.

Fiducial cuts were applied on the momenta and angles of the tracks in order to
select events from the well-understood regions of the detector. Included in these 
cuts was the removal of $13$ of the $288$ time-of-flight elements due to poor performance.
In addition, events where the missing $\pi^{0}$ was moving along the beam line,
$\cos{\theta_{c.m.}^{\pi^0}} > 0.99$, were cut in order to remove leakage from the 
$\gamma p \rightarrow p \pi^+ \pi^-$ reaction (because of the very forward angle, 
the center-of-mass and lab angles are very similar, so a $0.99$ cut in the center of mass
corresponds to an even tighter cut in the lab frame).  These arise from events in which 
two in-time photons occur where the higher-energy one is incorrectly associated with 
the charged tracks in the event.
The lower-energy photon causes the reaction $\gamma p\rightarrow p\pi^{+}\pi^{-}$.
When the event is reconstructed using the higher energy photon, there appears to 
be excess energy and missing momentum along the photon direction. In some cases,
this can be mis-reconstructed as a missing $\pi^{0}$ moving along the beam 
direction. Our cut at $\cos\theta_{c.m.}^{\pi^{0}} > 0.99$ removes these events. A more detailed 
description of the entire analysis procedure presented in this paper can be found 
in Ref.~\cite{williams-thesis}.

The resulting data have been sorted into bins in $W$. For the $\eta$, there are $76$ bins
from $W$ of $1.570$~GeV to $2.840$~GeV. From $1.570$~GeV to $2.100$~GeV, the bins
are $10$~MeV wide. From $W=2.100$~GeV to $2.360$~GeV, the bins are $20$~MeV wide.
In the range from $2.360$~GeV to $2.640$~GeV, the bins are $40$~MeV wide. Finally, there
is a $50$~MeV wide bin from $W=2.680$~GeV to $2.730$~GeV and a bin from $2.750$~GeV
to $2.840$~GeV. The $\eta^{\prime}$ data are divided into $44$ bins from $W=1.900$~GeV up
to $2.840$~GeV. The bin widths for the $\eta^{\prime}$ are the same as those used for the 
$\eta$.

\begin{figure}[h!]\centering
\subfigure[]{
  \includegraphics[width=0.4\textwidth]{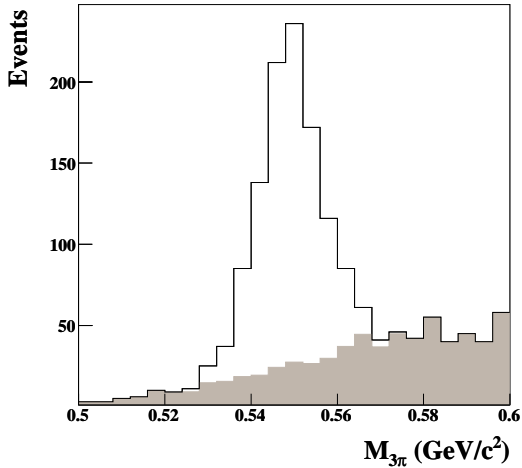}
}\\
\subfigure[]{
  \includegraphics[width=0.4\textwidth]{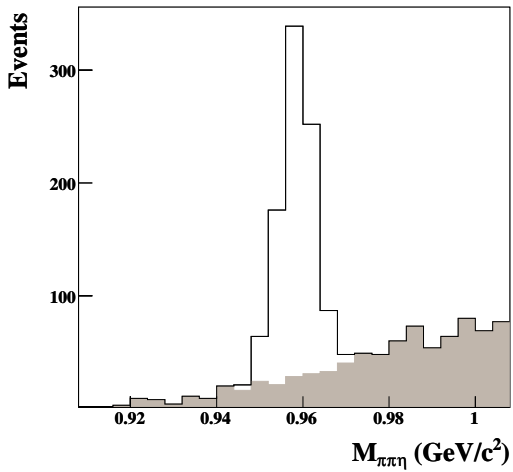}
}
\caption[]{\label{fig:sig-bkgd}
  Example bins demonstrating signal-background separation. The plots are 
  from the $W=2.11$~GeV, 
  $\cos{\theta_{c.m.}^{\eta(\eta^{\prime})}}=0.75$ kinematic bin for
  (a) $\eta$ and
  (b) $\eta^{\prime}$.
  The unshaded histograms are all of the data in each 4-MeV wide bin, while the shaded
  histograms are these same events weighted by the background factors, $1-Q$.
  See text for details.
}
\end{figure}

\section{\label{section:sig-bkgd}Signal-Background Separation}
In order to extract physical observables from the $\eta$ and $\eta^{\prime}$ 
photoproduction reactions studied here, it is necessary to separate background
events from signal events. While such separation could be done via a simple
side-band subtraction, our work on other reaction channels required the development
of a more sophisticated event-based procedure. This event-based method, described 
in detail elsewhere~\cite{williams-thesis,bkgd-preprint}, was used to separate 
signal and background events. 

The procedure takes advantage of the fact that in the invariant mass distributions,
the signal is a narrow structure, while the background is relatively featureless. 
The key feature of the procedure involves selecting each event's $n_c$ ``nearest 
neighbor'' events (we chose $n_{c} = 500$) using the quantity 
$\Delta\cos{\theta_{c.m.}^{\eta(\eta^{\prime})}}$ as a distance metric between events. 
Each subset of $n_{c}$ events occupies a very small region of phase space. Thus, 
the $M_{3\pi}(M_{\pi\pi\eta})$ distributions could safely be used to determine 
the probability of each event being a signal event---the event's $Q$-factor. 
The $M_{3\pi}$/$M_{\pi\pi\eta}$ distributions in each event's nearest neighbor event 
samples are fit to a narrow Gaussian (signal) plus a broad Gaussian and linear
(background) function to determine the $Q$-factors. A few example invariant mass distributions
are shown in Fig.~\ref{fig:sig-bkgd}.  Note that such a distribution and fit are generated 
for each event in order to determine the $Q$-factor for the event.
These \mbox{$Q$-factors} are then used to weight each event's contribution to the fits
that are in turn used to determine the detector acceptance. The $Q$-factors are also 
used to weight each event's contribution to the differential cross section.

Systematic studies were performed using different parametrizations of the
background (including up to fourth order polynomials). From these studies,
the systematic uncertainty in the yield extraction due to the choice of background 
shape is estimated to be $4.1$\%($3.1$\%) for the $\eta(\eta^{\prime})$ analysis.  The 
point-to-point uncertainties obtained from the individual fits varied depending 
on kinematics; however, it was typically about $4$\%--$5$\%. 

\section{\label{section:acc}Acceptance}
The typical method of computing a detector acceptance in a multi-particle final
state such as those studied here is an iterative procedure to achieve a 
\emph{physics model}. To determine if the model is good, one generates a 
Monte Carlo sample thrown according to the model. These events are then 
passed through an accurate detector simulation and the resulting event distributions
are then compared to the physics distributions for the signal events as measured
in the detector. After many iterations of the physics model, the two distributions 
should agree. At that point, one is able to use the physics model and the Monte 
Carlo sample to compute the detector acceptance. 

   There is an added complication to this procedure if there is background that 
is not easily separated from the signal. In such a case, one either needs to find 
a very clever way to effectively separate signal from background, or include the 
background events in the physics model. Generally, one is required to iterate the 
physics and background models until satisfactory agreement has been reached.

    In this analysis, we have taken a more systematic approach to computing the 
acceptance. First, we have used the $Q$-factors discussed earlier to allow us
to produce distributions of only signal events so that any model we use can 
ignore the background contribution. Second, we have used the fact that one can always 
expand any distribution in terms of partial waves. Thus our procedure involves
determining a partial wave expansion (our physics model), which will make the 
weighted Monte Carlo data agree with the observed signal sample. This partial 
wave expansion can be used to weight phase space generated events such that 
the Monte Carlo and data distributions agree. 

The efficiency of the detector was modeled using the standard CLAS GEANT-based 
simulation package and the Monte Carlo technique. A total of $100$ million $\eta$ 
and $80$ million $\eta^{\prime}$ events were generated pseudo-randomly, sampled from 
a phase space distribution. Each particle was propagated from the event vertex 
through the CLAS resulting in a simulated set of detector signals for each track. 
The simulated events were then processed using the same reconstruction software 
as the real data. In order to account for the event trigger used in this experiment 
(see Section~\ref{section:setup}), a study was performed to obtain the probability 
of a track satisfying the sector-based coincidences required by the trigger as a
function of kinematics and struck detector elements. The average effect of 
this correction in our analysis, which requires three detected particles, is
about $5$\%--$6$\%.

An additional momentum smearing algorithm was applied in order to better match 
the resolution of the Monte Carlo data to that of the real data. Its effects 
were studied using four-constraint kinematic fits performed on simulated 
$\gamma p \rightarrow p \pi^+ \pi^-$ events. After applying the momentum smearing 
algorithm, the same covariance matrix used for the real data also produced flat 
confidence level distributions in all kinematic regions for the Monte Carlo data 
as well. The simulated $\eta$ and $\eta^{\prime}$ events were then processed with 
the same analysis software as the real data, including the 1-constraint kinematic 
fits. At this stage, all detector and software efficiencies were accounted for. 

In order to evaluate the CLAS acceptance for the reactions 
${\gamma p \rightarrow p \eta}$ and ${\gamma p \rightarrow p\eta^{\prime}}$,
we chose to follow the same procedure that we used in obtaining the acceptance
for $\omega$ photoproduction~\cite{omega-prc}. In this procedure, we expand the 
scattering amplitude for the pseudoscalar photoproduction, $\mathcal{M}$, in a 
very large basis of $s$-channel waves as follows:
\begin{eqnarray}
  \label{eq:scattering-amp}
  \mathcal{M}_{m_i,m_{\gamma},m_f}(\cos{\theta_{c.m.}^{\eta(\eta^{\prime})}},
  \vec{\alpha})
  \hspace{0.25\textwidth}\nonumber \\ \approx
  \sum\limits_{J = \frac{1}{2}}^{\frac{11}{2}}\sum\limits_{P = \pm}
  \mathcal{A}_{m_i,m_{\gamma},m_f}^{J^P}(\cos{\theta_{c.m.}^{\eta(\eta^{\prime})}},
  \vec{\alpha}),
\end{eqnarray}
where $\vec{\alpha}$ denotes a vector of $34$ fit parameters, 
$m_i,m_{\gamma},m_f$ are the target proton, incident photon and final proton 
spin projections on the incident photon direction in the center-of-mass frame,
and $\mathcal{A}$ are the $s$-channel partial wave amplitudes. The $s$-channel 
structure of the amplitudes, along with the details concerning the fit parameters 
are described in Ref.~\cite{williams-thesis}. The amplitudes are evaluated using the 
\texttt{qft++} package~\cite{qft-paper}.

\begin{figure}\centering
\includegraphics[width=0.49\textwidth]{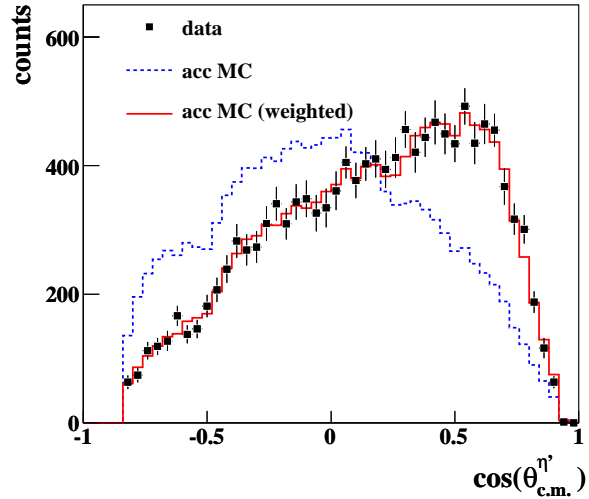}
\caption[]{\label{fig:acceptance}(Color Online) An example fit result in a typical
$W$ bin for the $\eta^{\prime}$ photoproduction. The data are shown as black squares,
the phase-space accepted Monte Carlo events are shown as the blue dashed line, while
the weighted Monte Carlo events discussed in Section~\ref{section:acc} are shown
as the solid red curve. The weighted Monte Carlo provides an excellent description of the
data.}
\end{figure}

Unbinned maximum likelihood fits were performed in each $W$ bin to obtain the estimators 
$\hat{\alpha}$ for the parameters $\vec{\alpha}$ in Eq.~(\ref{eq:scattering-amp}). 
Background events were removed using the $Q$-factors directly in the fit as discussed 
in Refs.~\cite{williams-thesis} and~\cite{bkgd-preprint}.
The results of these fits were used to obtain a physics weight, $I_{i}$, for each 
event. The weighted accepted Monte Carlo events fully reproduce the real data. An example
comparison is shown in Figure~\ref{fig:acceptance} for one $W$ bin of the $\eta^{\prime}$
data sample. The agreement between the weighted Monte Carlo and the data is very good.
We note here that the results of these fits are not interpreted as physics,
{\em i.e.} they are not considered evidence of resonance contributions to
$\eta(\eta^{\prime})$ photoproduction. They are simply used to provide a complete 
description of the data.

For a kinematic bin, the acceptance can be obtained as
\begin{equation}
  acc(W,\cos\theta_{\eta,\eta^{\prime}}) = \frac{\sum\limits_i^{N_{acc}} 
    I_i}{\sum\limits_j^{N_{gen}} I_j},
\end{equation}
where $N_{acc}(N_{gen})$ are the number of accepted (generated) Monte Carlo events in the 
bin and the $I$'s are the event weights discussed above. An accurate physics generator 
would use the factors of $I$ during the event generation stage, rather than weighting 
the events. The resulting acceptance calculation would be the same, modulo statistical 
fluctuations.

\section{\label{section:norm}Normalization}
The measured rate of electrons detected by the tagger was used to compute
the number of photons incident on the target by sampling tagger hits not in
coincidence with the CLAS. These rates were integrated over the
live-time of the experiment to obtain the total photon flux associated with
each tagger element. Losses of photons in the beam line due to effects such
as beam collimation were determined during dedicated runs using a total-absorption 
counter placed downstream of the CLAS~\cite{gflux}.

The main method to calculate the experimental live-time during the ``g11a''
run was based on a clock. This live-time calculation was checked by using the 
counts of a Faraday cup located downstream of the detector. While the Faraday
cup is a standard device for electron beam intensity measurements, in this case
it was counting many fewer events produced by interactions of the photon beam 
with the target.  Consequently, the statistical error of this second live-time 
measurement was high. However, the Faraday-cup-based measurement
allowed us to observe that at maximum electron beam current, the actual dead
time was about a factor of two higher than that given by the clock-based measurement.
The high statistical error of the Faraday-cup-based measurement led to an
uncertainty in the absolute live-time measurement of about $3$\%.

As discussed in Section~\ref{section:setup}, tagger counters $1$-$40$ were in the trigger, 
while counters $41$-$61$ were not. In order to have an event originating from 
a photon in the ``untriggered'' part of the tagger, the detector needed to record
a random hit in one of the ``triggered'' counters during the trigger time window. 
The probability of such an occurrence can be calculated using the electron rates. 
For the data here, we found that the probability of this happening is $46.7$\%.
This factor is then used to scale the flux in the untriggered counters.

Events in the $W=~1.955$~GeV bin span the boundary between tagger counters $40$ and
$41$. Because events in this energy range arose from both triggered and untriggered 
counters, the flux in this bin was deemed unreliable. Thus, we report no cross sections 
for this energy.  In addition to the above bin, the electronics in one of the tagger 
channels was not working properly during the run. This led to inaccurate flux 
measurements in the energy range $W~=~2.73-2.75$~GeV. Differential cross sections 
will not be reported at these energies as well.

\section{\label{section:errors}Systematic Uncertainties}
An overall acceptance uncertainty of $5$\%--$7$\%, depending on center-of-mass 
energy, was estimated for this analysis in Ref.~\cite{omega-prc}. This includes
uncertainties due to particle identification ($1.3$\%) and kinematic fitting confidence 
level cuts ($3$\%), along with a relative acceptance uncertainty estimated by studying
the agreement of physical observables obtained from each of the independent 
CLAS sectors ($4-6$\%). The uncertainty on the normalization calculation was also
estimated in Ref.~\cite{omega-prc} and found to be $7.9$\%. This includes 
contributions from photon transmission efficiency and live-time calculations.

The acceptance and normalization uncertainties discussed above were then
combined with contributions from target density and length ($0.2$\%), along with 
branching fraction ($0.4$\% for the $\eta$, $1.5$\% for the $\eta^{\prime}$) 
to obtain a total uncertainty, excluding the point-to-point contributions from 
signal-background separation (from the fits), of about $9$\%--$11$\%. 
The additional $4.1$\% and $3.1$\% global signal-background uncertainties 
discussed in Section~\ref{section:sig-bkgd} must then be added in quadrature 
giving total uncertainties of about $10$\%--$12$\% for both the $\eta$ and $\eta^{\prime}$.
These errors are summarized in Table~\ref{tab:systematics}.
\begin{table}[h!]\centering
\begin{tabular}{lrr} \hline
Error & $\eta$ & $\eta^{\prime}$ \\ \hline
$\Delta$ TOF Cut (PID) & $1.3$\% & $1.3$\% \\
Confidence Level Cut   & $3$\%   & $3$\% \\
Relative Acceptance             & $4$\%--$6$\% & $4$\%--$6$\% \\
Normalization          & $7.9$\% & $7.9$\% \\
Target Length          & $0.2$\% & $0.2$\% \\
Branching Fraction     & $0.4$\% & $1.5$\% \\ 
Background Shape       & $4.1$\% & $3.1$\% \\ \hline
Total                  & \multicolumn{2}{c}{$10$\%--$12$\%} \\ \hline
\end{tabular}
\caption[]{\label{tab:systematics}A summary of the systematic errors associated
with these measurements. The total presents the range of values seen across all
data bins.}
\end{table}
\begin{figure*}[h!]
  \centering
  \includegraphics[width=1.0\textwidth]{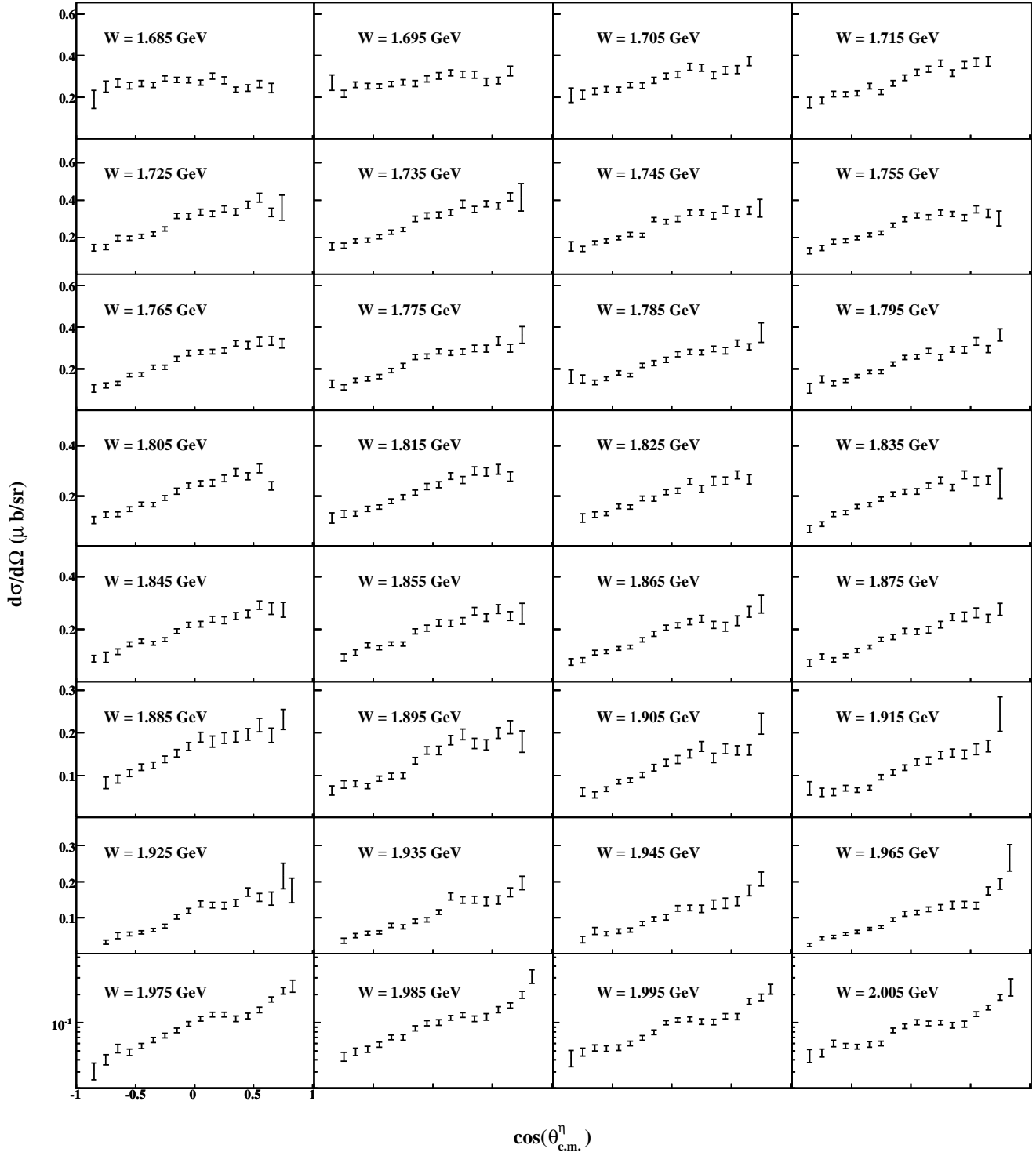}
\caption[]{\label{fig:dsigma-eta-1}
  $\frac{d\sigma}{d\Omega}(\mu$b/sr) {\em vs.} $\cos{\theta^{\eta}_{c.m.}}$ for the
  $\gamma p \rightarrow p \eta$ reaction. Note that the vertical axis is linear
for $W$ up to $1.965$~GeV and logarithmic above that.}
\end{figure*}

\begin{figure*}[h!]
  \centering
  \includegraphics[width=1.0\textwidth]{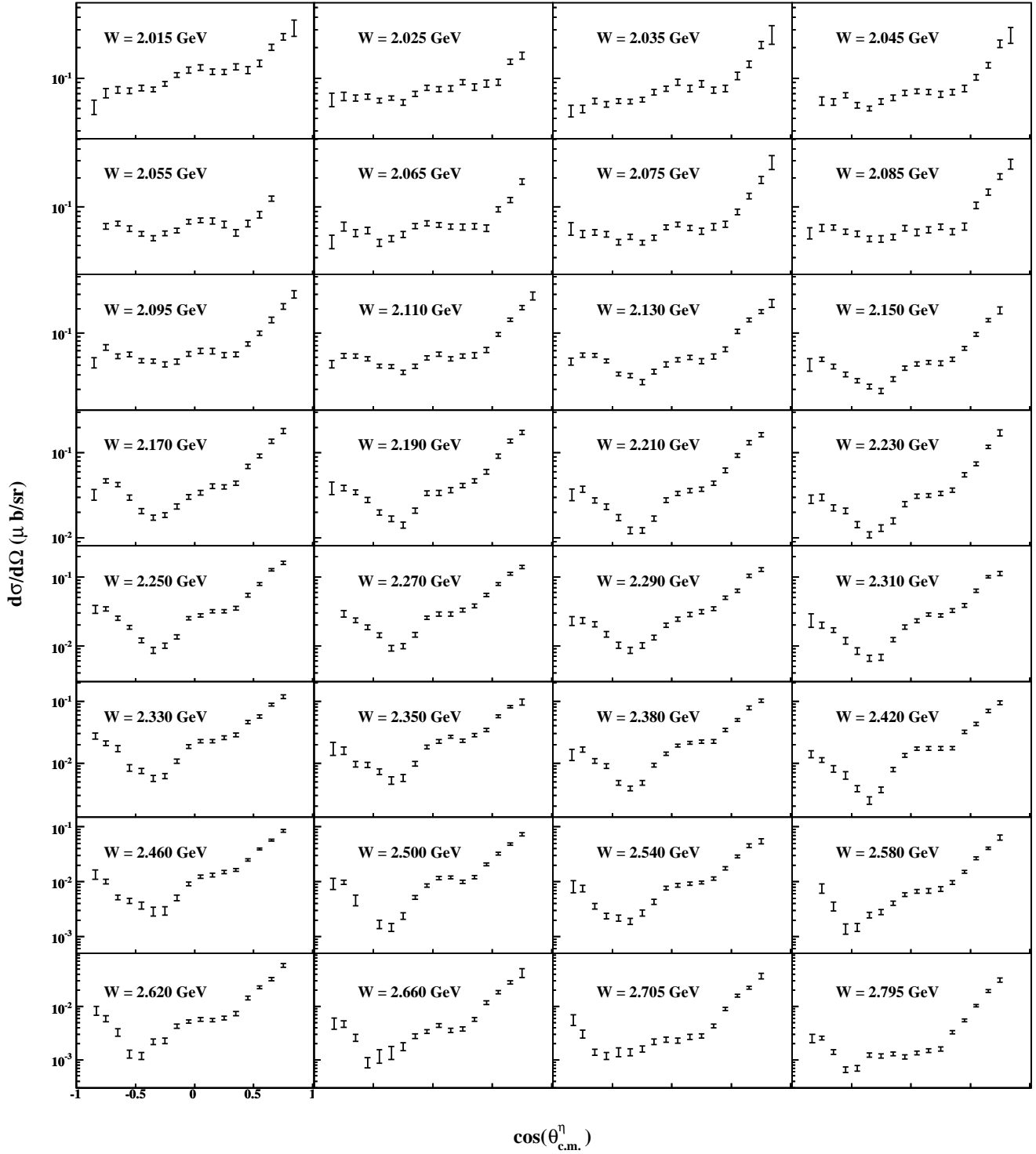}
\caption[]{\label{fig:dsigma-eta-2}
  $\frac{d\sigma}{d\Omega}(\mu$b/sr) {\em vs.} $\cos{\theta^{\eta}_{c.m.}}$ for the
  $\gamma p \rightarrow p \eta$ reaction.  Note the logarithmic scale on the vertical
axis.}
\end{figure*}

\begin{figure*}[h!]
  \centering
\includegraphics[width=1.0\textwidth]{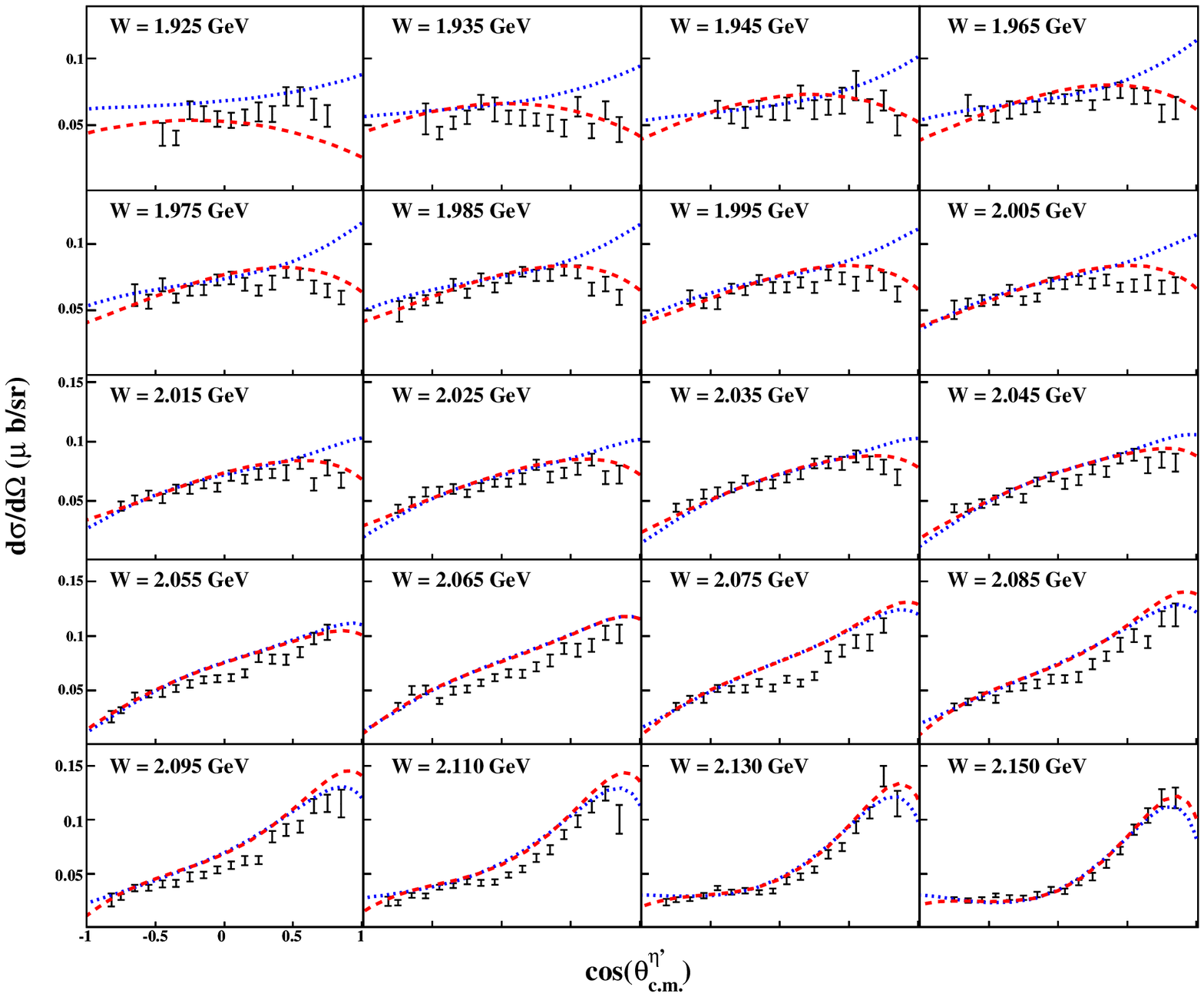}
\caption[]{\label{fig:dsigma-etaprime}
  $\frac{d\sigma}{d\Omega}(\mu$b/sr) {\em vs.} $\cos{\theta^{\eta^{\prime}}_{c.m.}}$ for the
  $\gamma p \rightarrow p \eta^{\prime}$ reaction. Note that the vertical axis is linear.
The (red) dashed line and (blue) dotted line are the results from Tabs.~II  and IV of 
Ref.~\cite{kanzo}, respectively.}
\end{figure*}

\begin{figure*}[h!]
  \centering
\includegraphics[width=1.0\textwidth]{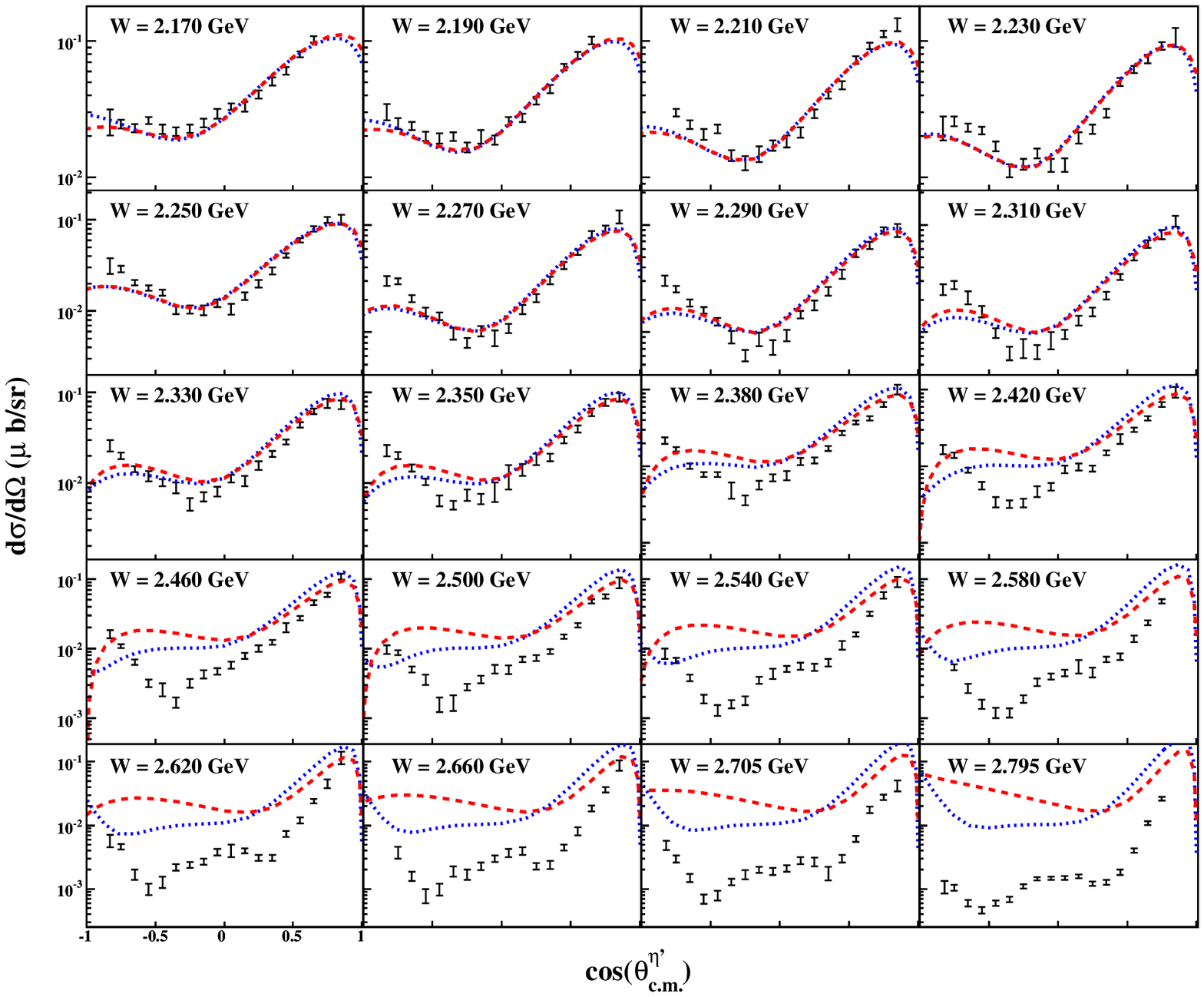}
\caption[]{\label{fig:dsigma-etaprime-2}
  $\frac{d\sigma}{d\Omega}(\mu$b/sr) {\em vs.} $\cos{\theta^{\eta^{\prime}}_{c.m.}}$ for the
  $\gamma p \rightarrow p \eta^{\prime}$ reaction. Note that the vertical axis is logarithmic.
  The (red) dashed line and (blue) dotted line are the results from Tabs.~II and IV of 
  Ref.~\cite{kanzo}, respectively.}
\end{figure*}

\section{\label{section:results}Results}

For the differential cross section calculations,  $d\sigma/d\Omega$, each 
center-of-mass energy bin was divided into $20$ bins in $\cos{\theta_{c.m.}^{\eta(\eta^{\prime})}}$ 
of width $0.1$; however, results could not be extracted in every bin due to limitations 
in the detector acceptance. In total, $1082$ $\eta$ and $682$ $\eta^{\prime}$ cross section 
points are reported. The centroid of each bin is reported as the mean of the range of the bin
with non-zero acceptance. The results are shown in Figs.~\ref{fig:dsigma-eta-1}, 
\ref{fig:dsigma-eta-2}, \ref{fig:dsigma-etaprime} and \ref{fig:dsigma-etaprime-2} 
and are available online in
electronic form~\cite{clasdb}. The error bars contain the 
uncertainties in the signal yield extraction (point-to-point signal-background 
separation uncertainties and statistical uncertainties in the number of signal events), 
along with statistical uncertainties from the Monte Carlo acceptance calculations. The 
global systematic uncertainties, discussed in Section~\ref{section:errors}, are 
estimated to be between $10$\%-$12$\%, depending on center-of-mass energy.

Both data sets have a forward peak that becomes more prominent with increasing
energy, most likely due to some $t$-channel exchange mechanism. Both data sets
also have a backward peak at our highest energies, this could be indicative
of $u$-channel contributions. There are other interesting features in the data sets
which might be evidence for resonance production; however, a partial wave analysis 
would need to be performed to determine if this is the case. In the $\eta$ photoproduction,
the most prominent of these are a bump near $\cos\theta=0.2$ in the low-energy data and
a dip and a shoulder near $\cos\theta=-0.1$ in the $2.1$ to $2.6$~GeV region. For the
$\eta^{\prime}$ there is a similar dip and shoulder.  

\begin{figure*}[t!]
  \centering
  \includegraphics[width=0.95\textwidth]{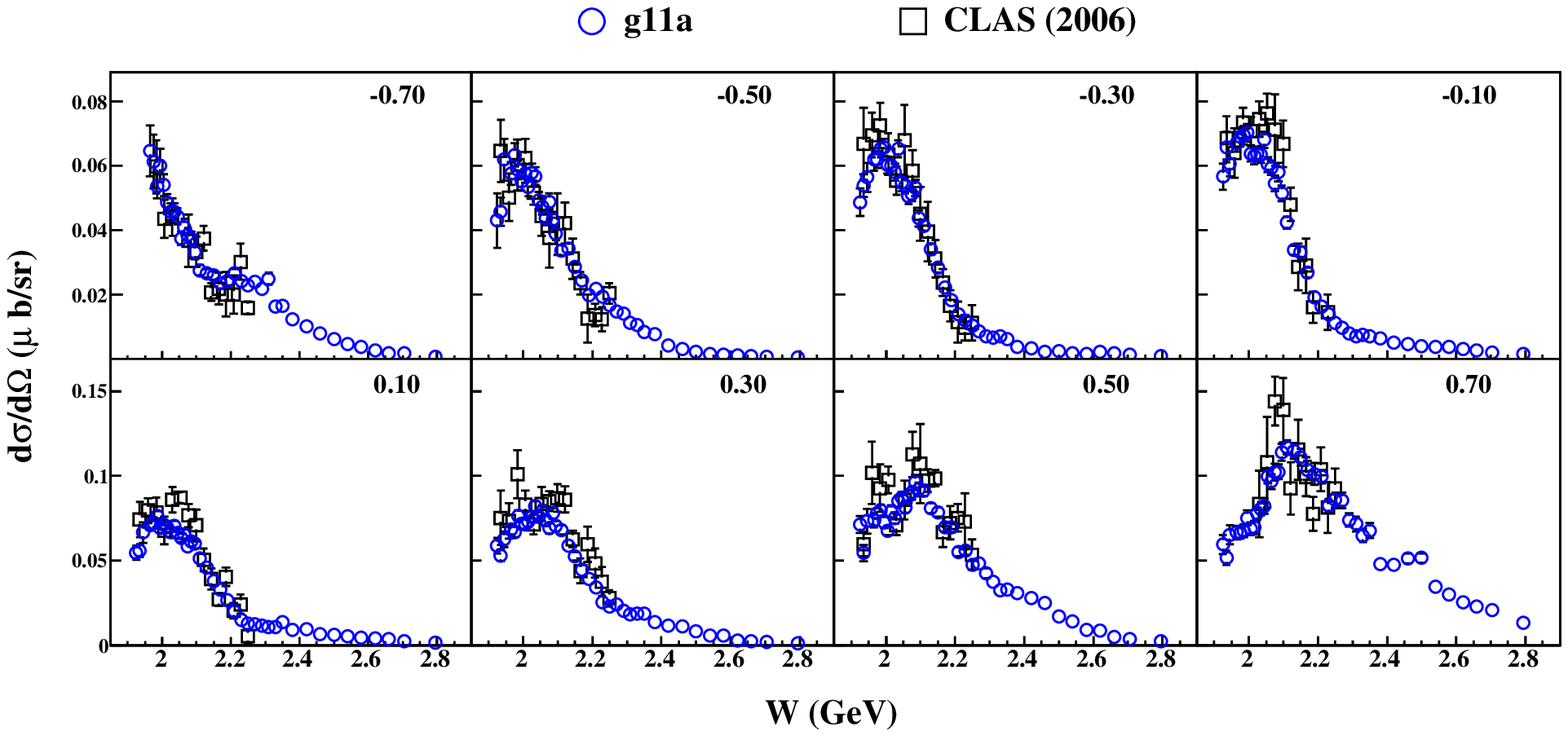}
\caption[]{\label{fig:compare-etaprime}
  (Color Online)
  $\frac{d\sigma}{d\Omega}(\mu$b/sr) {\em vs.} $W$ (GeV) for the
  $\gamma p \rightarrow p \eta^{\prime}$ reaction,   in 
  $\cos{\theta_{c.m.}^{\eta^{\prime}}}$ bins, from previous CLAS 
  data~\cite{dugger-etaprime} (open squares) and this work (blue open circles). 
  For this comparison, our results were merged into $10$ $\cos{\theta_{c.m.}^{\eta^{\prime}}}$ 
  bins, eight of which exactly overlap the earlier CLAS measurements shown here (see text 
  for details). The centroid in $\cos\theta^{\eta^{\prime}}_{c.m.}$ of each bin is labeled on the plot.
}
\end{figure*}

Figure~\ref{fig:compare-etaprime} shows a comparison of our $\eta^{\prime}$ results to 
previously published CLAS data~\cite{dugger-etaprime}. The previous CLAS results 
were published using $\cos{\theta_{c.m.}^{\eta^{\prime}}}$ bins of width 0.2, as opposed to 
the $0.1$ width bins used in our work. To make this comparison, we have merged our 
bins (using a weighted average) to obtain the same binning as the previous CLAS results.
The agreement is good with most data points agreeing within the error bars.

\clearpage


\begin{figure*}[h!]
  \centering
  \includegraphics[width=0.99\textwidth]{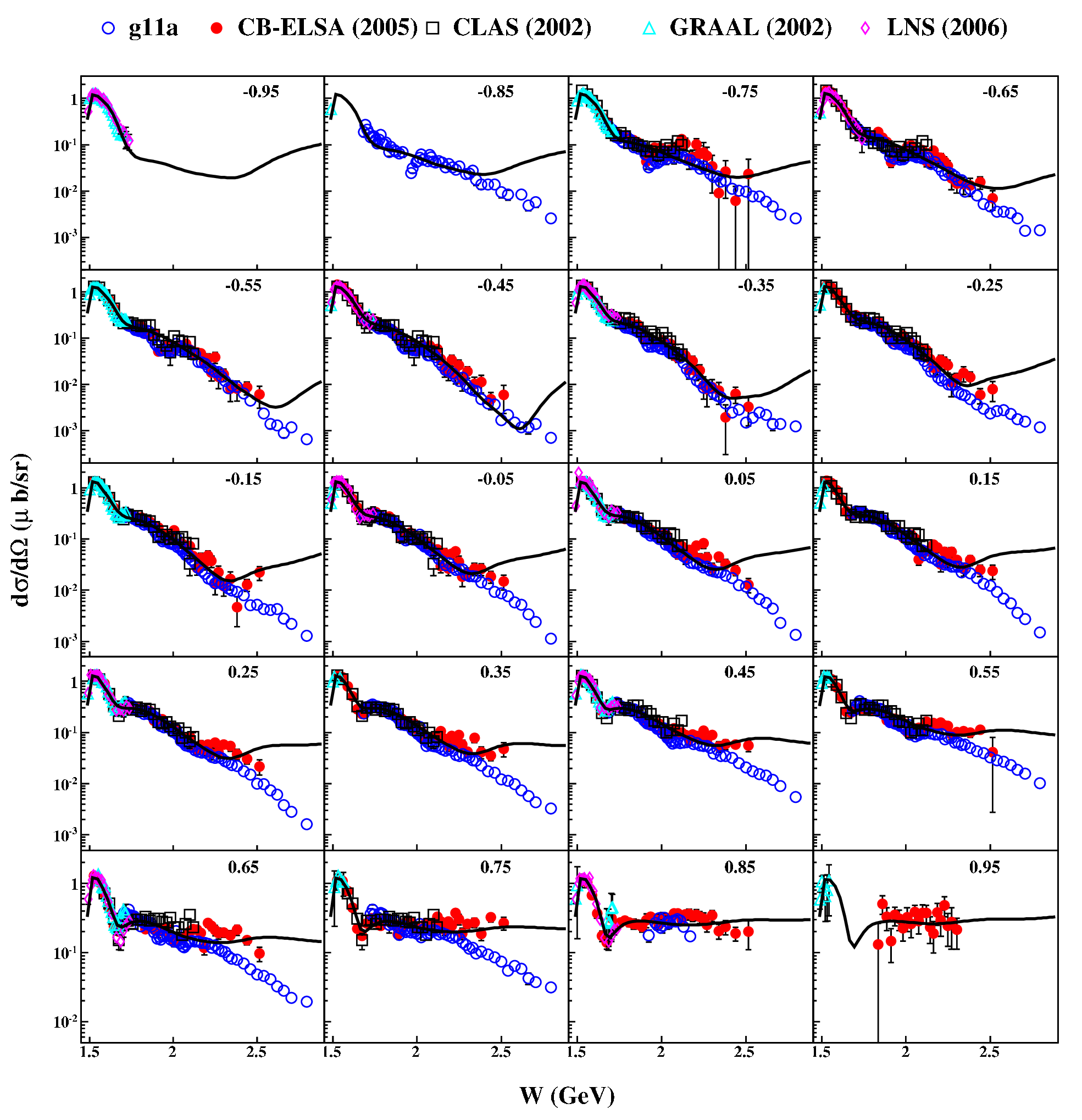}
\caption[]{\label{fig:compare-eta}
  (Color Online)
  $\frac{d\sigma}{d\Omega}(\mu$b/sr) {\em vs.} $W$ (GeV) for the
  $\gamma p \rightarrow p \eta$ reaction, in $\cos{\theta_{c.m.}^{\eta}}$ bins,
  from CB-ELSA~\cite{crede}(red filled circles),  previous CLAS 
  data~\cite{dugger-eta}(open squares), GRAAL~\cite{graal02}(light-blue open triangles),
  LNS~\cite{lns06}(purple open diamonds)  and this work (blue open circles). 
  The centroid of each bin in $\cos\theta^{\eta}_{c.m.}$ is labeled on each plot.
  The solid line is a SAID~\cite{said} fit to the earlier data.}
\end{figure*}

Figures~\ref{fig:dsigma-etaprime} and~\ref{fig:dsigma-etaprime-2} also shows the results from Tabs.~II 
and IV of Ref.~\cite{kanzo} (a relativistic meson-exchange model which includes
various resonance contributions).  Five versions of the model are presented 
in \cite{kanzo}; each fit to the previously published CLAS 
data~\cite{dugger-etaprime}. The previous data
were not able to distinguish between the five versions of the model.  
Near threhold, the Tab.~II results clearly provide a better description of our 
data than those of Tab.~IV. 
Given the poor performance of the models at higher energies,
{\em i.e.} energies greater than what the models were fit to ($W>2.25$~GeV), 
physics claims cannot be made from these models at this time. 
Our data, however, provide additional angular and energy coverage, 
along with increased precision, which should allow for more reliable 
extraction of resonance contributions to $\eta'$ photoproduction by providing 
more stringent constraints on the models.

Figure~\ref{fig:compare-eta} shows a comparison of our $\eta$ results to published
results from CLAS~\cite{dugger-eta} and CB-ELSA~\cite{crede}. The agreement between 
the three sets of measurements is fair to good at lower energies; however, at higher 
energies and forward angles large discrepancies begin to develop. The highest energy 
results reported by CB-ELSA are approximately four times larger than our measurements 
in the most forward angles. Even for $\cos{\theta_{c.m.}^{\eta}} \approx 0$, the
CB-ELSA results are approximately two times larger than ours at these energies.
The figure also shows a SAID fit~\cite{said} to the earlier data. Given the differences 
between our measurement and the earlier data, the poor agreement between the 
fit and our results is not surprising. 
Including our results into the fit will likely have a significant impact on the
extracted physics.

\begin{figure}[h!]\centering
\includegraphics[width=0.49\textwidth]{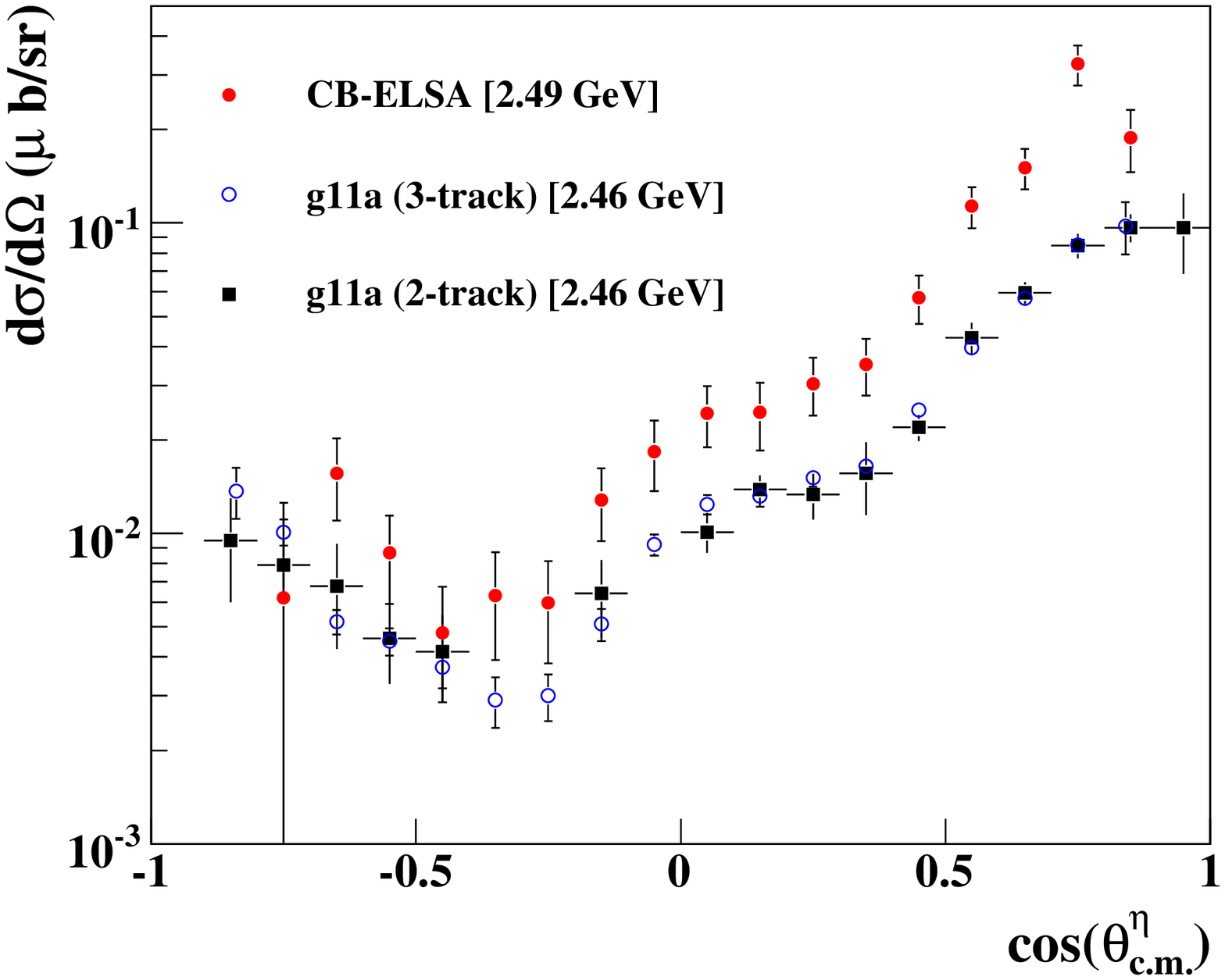}
\caption[]{\label{fig:compare}
(Color Online) $\frac{d\sigma}{d\Omega}$ ($\mu$b/sr) vs. $\cos{\theta ^{\eta}_{c.m.}}$: 
  for the reaction $\gamma p\rightarrow p\eta$.
  Differential cross sections near $W = 2.46$~GeV from CB-ELSA~\cite{crede}
  (red filled circles) and from CLAS. The CLAS {\em g11a} results presented in this 
  work, which required detection of $p\pi^+\pi^-$ (blue open circles), and the ``two-track'' 
  results discussed in the text, which only required the $p\pi^+$ to be detected 
  (black filled squares) are in excellent agreement. The background levels in the 
  ``two-track'' topology were too high to permit  a reliable extraction of the signal 
  yield for $-0.4 < \cos{\theta^{\eta}_{c.m.}} < -0.2$. 
  The ``two-track'' error bars are purely statistical and do not contain any 
  systematic uncertainty estimates on the signal-background separation.
}
\end{figure}

This extremely large disagreement at higher energies and forward angles motivated 
us to extract the $\gamma p \rightarrow p \eta$
cross sections using an alternate procedure. This was carried out
for points in our $W=2.46$~GeV bin (which overlaps the second highest CB-ELSA bin).
In this alternate procedure, we required the detection of only the $p\pi^+$, while 
ignoring the $\pi^{-}$. The $\eta$ signal was then identified in the missing mass
off the proton.  This topology has both a very different acceptance and a significantly
larger background than the one in which all three charged particles are detected; 
however, the results obtained for the two topologies were in excellent 
agreement (see Figure~\ref{fig:compare}). This study, along with the fact that our results
 from this data set in several other channels (see Ref.~\cite{omega-prc} 
and~\cite{mccracken}) are in good agreement with the world's data, has led us to 
conclude that it is very unlikely that  there is some unknown acceptance or normalization 
issue present in our data.

We can offer no explanation as to why the CB-ELSA results differ so much from 
ours; however, the self-consistency of the results obtained from our data set,
using two distinct topologies, along with the high level of agreement with the 
world's data of cross sections extracted for other reactions from this same data set 
provide a high level of confidence in the results presented in this paper.

Ultimately, one expects that these data, combined with other measurements, will
facilitate a large scale partial wave analysis that will be able to identify the 
baryon resonance contributions to these cross sections. While we did attempt to 
carry out single-channel partial wave analyses of both of these channels~\cite{krahn} (similar 
to that in Ref.~\cite{omega-prd}), the limited
number of observables prevented us from drawing clear conclusions. Together
with new measurements involving polarized beams and targets, these results should
enable a deeper understanding of the nucleon resonances in the future.

\section{\label{section:conclusions}Conclusions}

In summary, experimental results for $\eta$ and $\eta^{\prime}$ photoproduction 
from the proton have been presented in the energy regime from near threshold up to 
$W = 2.84$~GeV. A total of $1082$ $\eta$ and $682$ $\eta^{\prime}$ cross section 
points are reported. The $\eta^{\prime}$ results are the most precise to-date and 
provide the largest energy and angular coverage. The $\eta$ measurements extend the 
energy range of the world's large-angle results by approximately $300$~MeV.
Unfortunately, discrepancies exist between the $\eta$ results presented here and those
previously published by CB-ELSA~\cite{crede} at higher energies. We look forward
to seeing the impact these new results will have on existing models of baryon photoproduction.

\begin{acknowledgments}
We thank Megan Friend for her valuable work on background subtraction studies
for these reactions. We also thank the staff of the Accelerator and the Physics Divisions at
Thomas Jefferson National Accelerator Facility who made this experiment possible.  
This work was supported in part by the U.S. Department of Energy
(under grant No. DE-FG02-87ER40315), the National Science Foundation,
the Italian Istituto Nazionale di Fisica Nucleare, 
the French Centre National de la Recherche Scientifique, 
the French Commissariat \`{a} l'Energie Atomique, 
the Science and Technology Facilities Council (STFC),
and the Korean Science and Engineering Foundation.  
The Southeastern Universities Research Association (SURA)
operated Jefferson Lab under United States DOE contract
DE-AC05-84ER40150 during this work.
\end{acknowledgments}


\begin{thebibliography}{99}
\bibitem{clasdb}
Electronic versions of the cross section data presented in this paper can
be found online at \texttt{clasweb.jlab.org/physicsdb}  .
\end{thebibliography}
\end{document}